\documentclass [%
superscriptaddress,
reprint,
 amsmath,amssymb,
 showkeys
]{revtex4-1}

\usepackage{graphicx}
\usepackage{dcolumn}
\usepackage{bm}
\usepackage{color}
\usepackage{float}
\usepackage[utf8]{inputenc}
\usepackage[mathlines]{lineno}
\usepackage{multirow}
\usepackage{hyperref}
\usepackage{amsmath}
\usepackage{xr}
\usepackage{soul,xcolor}
\soulregister\cite7
\soulregister\ref7
\setstcolor{blue}
\newcommand{\MABr}{CH$_3$NH$_3$PbBr$_3$}
\newcommand{\Cs}{CsPbBr$_3$}

\newcommand{\MABrws}{CH$_3$NH$_3$PbBr$_3$ }
\newcommand{\Csws}{CsPbBr$_3$ }

\newcommand{\MAws}{[CH$_3$NH$_3$]$^+$ }

\newcommand{\wn}{cm$^{-1}$}
\newcommand{\wnws}{cm$^{-1}$ }

\begin{document}

\title{Local polar fluctuations in lead halide perovskite crystals}

\author{Omer Yaffe}
\thanks{These authors contributed equally}
\affiliation{Department of Chemistry, Columbia University, New York, NY 10027, USA}
\author{Yinsheng Guo}
\thanks{These authors contributed equally}
\affiliation{Department of Chemistry, Columbia University, New York, NY 10027, USA}
\author{Liang Z. Tan}
\affiliation{Department of Chemistry, University of Pennsylvania, Philadelphia, Pennsylvania 19104, USA}
\author{David A. Egger}
\affiliation{Department of Materials and Interfaces, Weizmann Institute of Science, Rehovoth 76100, Israel}
\author{Trevor Hull}
\affiliation{Department of Chemistry, Columbia University, New York, NY 10027, USA}
\author{Constantinos C. Stoumpos}
\affiliation{Materials Science Division, Argonne National Laboratory, Argonne, Illinois 60439, USA}
\author{Fan Zheng}
\affiliation{Department of Chemistry, University of Pennsylvania, Philadelphia, Pennsylvania 19104, USA}
\author{Tony F. Heinz}
\affiliation{Department of Applied Physics, Stanford University, Stanford, CA 94305, USA}
\affiliation{SLAC National Accelerator Laboratory, Menlo Park, California 94025, United States}
\author{Leeor Kronik}
\affiliation{Department of Materials and Interfaces, Weizmann Institute of Science, Rehovoth 76100, Israel}
\author{Mercouri G. Kanatzidis}
\affiliation{Materials Science Division, Argonne National Laboratory, Argonne, Illinois 60439, USA}
\affiliation{Department of Chemistry, Northwestern University, Evanston, Illinois 60208, USA} 
\author{Jonathan S Owen}
\affiliation{Department of Chemistry, Columbia University, New York, NY 10027, USA}
\author{Andrew M. Rappe}
\affiliation{Department of Chemistry, University of Pennsylvania, Philadelphia, Pennsylvania 19104, USA}
\author{Marcos A. Pimenta}
\affiliation{Department of Chemistry, Columbia University, New York, NY 10027, USA}
\affiliation{Departamento de Fisica, Universidade Federal de Minas Gerais, 30123-970 Belo Horizonte, Brazil}
\author{Louis E. Brus}
\email{leb26@columbia.edu}
\affiliation{Department of Chemistry, Columbia University, New York, NY 10027, USA}

\date{\today}

\begin{abstract}

Hybrid lead-halide perovskites have emerged as an excellent class of photovoltaic materials. 
Recent reports suggest that the organic molecular cation is responsible for local polar fluctuations that inhibit carrier recombination.  
We combine low frequency Raman scattering with first-principles molecular dynamics (MD) to study the fundamental nature of these local polar fluctuations. Our observations of a strong central peak in the cubic phase of both hybrid (\MABr) and all-inorganic (\Cs) lead-halide perovskites show that anharmonic, local polar fluctuations are intrinsic to the general lead-halide perovskite structure, and not unique to the dipolar organic cation. 
MD simulations indicate that head-to-head Cs motion coupled to Br face expansion, occurring on a few hundred femtosecond time scale, drives the local polar fluctuations in CsPbBr$_3$.
\end{abstract}
%

\keywords{Local polar fluctuations; Hybrid perovskites; Low frequency Raman scattering; molecular dynamics; radiation detection; {Central peak}; }
\maketitle

Lead-halide hybrid perovskite crystals have emerged as promising materials for inexpensive and efficient solar cells. The rapid increase in solar cell  power conversion efficiency\cite{Gratzel2014,Zhou2014} provides impetus to gain fundamental understanding of the interplay between their electronic and structural properties.   

Recent reports show that even without defects, hybrid perovskite crystals display significant structural fluctuations~\cite{Worhatch08p1272,Mosconi2014,Carignano15p8991,Goehry15p19674,Quarti2015,Even16p6222,Leguy15p7124,Weller2015,Frost16p528,Egger16p573,Neukirch2016,Motta2015}. The organic molecular cation plays an important role in the structural fluctuations of the hybrid perovskites, as supported by dielectric measurements, neutron scattering, and molecular dynamics~\cite{Poglitsch1987,Bakulin2015,Mattoni2015,Mattoni16p529}. These studies revealed molecular dipole orientations fluctuating on a picosecond time scale, with the octahedral halide cages exhibiting  large anharmonic thermal fluctuations in response.~\cite{Chung12p8579,DaSilva2015} These large thermal polar fluctuations may result in local dynamic fluctuations of the electronic band gap and thus play a key role in determining electronic properties of hybrid crystals~\cite{Mosconi2014,Motta2015,Quarti2016}. 

Here, we investigate the nature of this unique structural behavior by combining low-frequency Raman scattering with MD simulations. 
Our experimental approach is to conduct a comparative study of CH$_3$NH$_3$PbBr$_3$, a hybrid perovskite, and \Cs, an all-inorganic halide perovskite with a similar band gap and structural phase sequence.\cite{Stoumpos2013a,Xing2014} We show that anharmonic effects~\cite{Mattoni16p529} are not limited to hybrid perovskites with molecular cations, but are also present in the  all-inorganic \Cs. In both CH$_3$NH$_3$PbBr$_3$ and CsPbBr$_3$, we observe a ``central" (zero-frequency) peak in their Raman spectra, which grows in prominence from a small background signal in the low temperature orthorhombic phases to the dominant spectral feature in the cubic phases. Such a central peak is a signature of local polar thermal fluctuations ~\cite{Diantonio93p5629,Farhi99p599,Vugmeister06p174117}. Raman central peaks do not arise in purely harmonic systems, which have well-defined Raman lines at finite frequencies. Instead, Raman central peaks are associated with polar correlations occurring over a range of time scales, which give rise to spectral weight over a continuum of frequencies. The presence of the central peak in both CH$_3$NH$_3$PbBr$_3$ and CsPbBr$_3$ shows that local polar fluctuations exist in both systems, even when the polar molecular cation CH$_3$NH$_3$$^+$ is replaced by the atomic cation Cs$^+$. Using first-principles MD simulations, we assign the central peak in CsPbBr$_3$ to anharmonic, head-to-head motion of Cs cations in the cubo-octahedral voids of the perovskite structure combined with the distortion of PbBr$_6$ octahedra. These local polar fluctuations occur on the hundred femtosecond time scale.

We performed Raman scattering measurements on high quality \MABrws and \Csws single crystals over a range of temperatures, thus probing for the orthorhombic, tetragonal, and cubic phases. Raman scattering spectra were obtained with excitation by 1.96 eV CW HeNe laser. By using this excitation energy, we benefit from a pre-resonance Raman effect where the scattering cross section is enhanced via proximity to the band gap optical transition.\cite{Loudon1963} Along with our Raman experiments, we compute theoretical Raman spectra of \mbox{\Csws} from first-principles MD,  based on density functional theory (DFT). These calculations were performed at 80~K and 500~K, using lattice parameters of the orthorhombic and cubic \mbox{\Csws}unit cell~\cite{Stoumpos2013a}, respectively, which were optimized with dispersion-corrected DFT ~\cite{Thomas2013}(see supplemental material). We calculate semiclassical Raman spectra from the autocorrelation function of the polarizability obtained from first-principles~\cite{Thomas2013}.  In boththe experimental and calculated Raman spectra, we observe a series of sharp peaks superimposed on a broad central peak exhibiting a highly temperature-dependent intensity. This agreement between spectroscopy and theory allows us to extract and identify the atomic motions responsible for the central peak in the Raman spectra.

\begin{figure}
\centering
\includegraphics[scale=0.6]{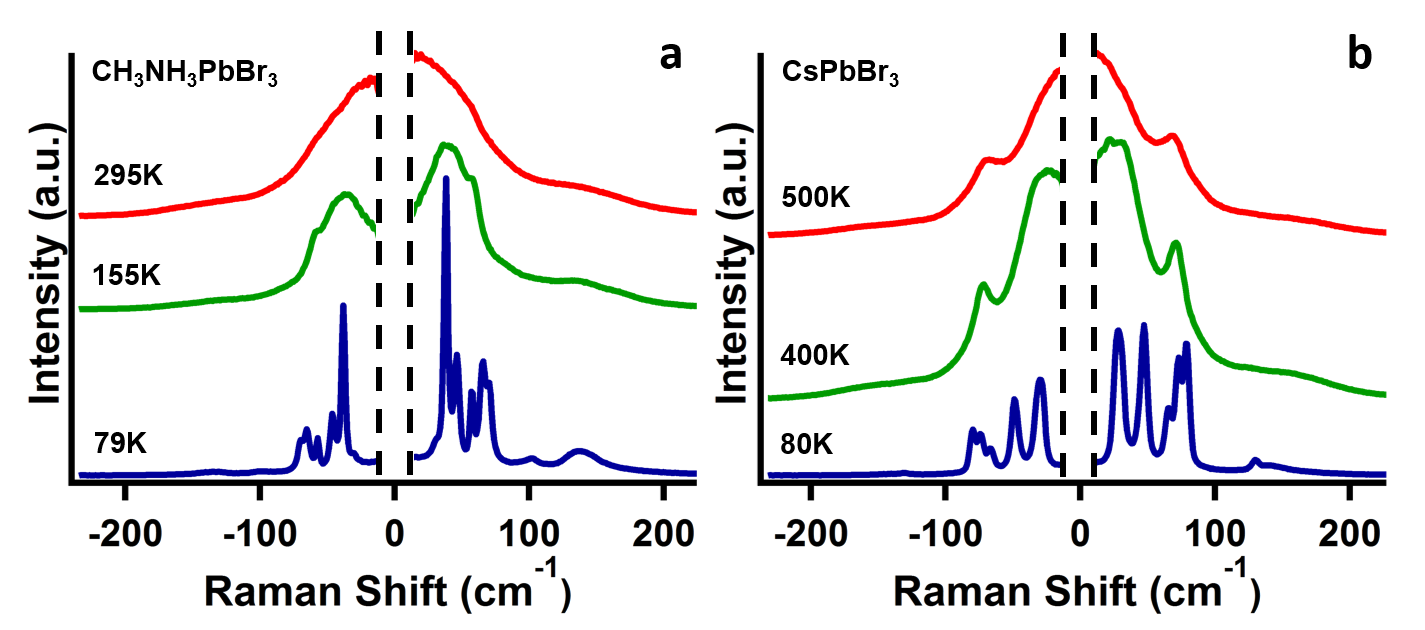}
\caption{
Low-frequency Raman spectra of hybrid and inorganic lead-halide perovskite crystals. The spectral region obscured by the notch-filter, between $\pm$10 \wnws (marked by the vertical dashed lines), has been deleted. (a) \MABrws and (b) \Csws in the orthorhombic phase (blue), tetragonal phase (green), and cubic phase (red), showing growth of central peak with temperature in both materials.
}
\label{fig:threephases}
\end{figure}

In figure~\ref{fig:threephases} we compare the Raman scattering spectra of \MABrws and \Csws single crystals across their similar phase sequence (see supplemental material for x-ray diffraction data on phase sequence vs temperature for \MABrws\cite{Poglitsch1987,Mashiyama2007} and \Csws\cite{Stoumpos2013a,Hirotsu1974}). Note that the transition temperatures of \Csws are much higher than those of \MABr.

 At 80~K, sharp, well-resolved Raman spectra are observed for \MABrws and \Csws crystals (bottom-blue curves in Figures ~\ref{fig:threephases}a and b, respectively). Both spectra share similar features: two strong modes below 50~\wn, a set of three peaks around 70--80~\wn, and a broad, low-intensity feature centered around 135~\wn. 

Unlike the low-temperature spectra, the Raman spectra in the tetragonal (Figure \ref{fig:threephases}a,b, middle-green) and cubic (Figure \ref{fig:threephases}a,b, top-red) phases  are diffuse, composed of a central peak with a significant spectral continuum underlying broad Raman transitions at positions corresponding to modes modestly softened from the sharpe 70~\wnws and 120~\wnws features presented in the lower temperature orthorhombic spectra. {Using the Debye relaxation model~\cite{Bouziane2003}, we find that time scale corresponding to the shape and width of the central peak, is a few hundreds femtoseconds for both crystals. Details of modeling the scattering data are presented in the supplemental material.}

XRD studies show that both crystals are cubic in the high temperature phase ~\cite{Poglitsch1987,Stoumpos2013a}. Factor group analysis of the averaged, cubic structure predicts no Raman activity~\cite{Maalej1997}. The fact that we observe diffuse Raman spectra and the central peak suggests that these structures are highly dynamic, fluctuating among different non-cubic structures in such a way as to appear cubic on average~\cite{Quarti2016}.

\begin{figure}
\centering
\includegraphics[scale=0.6]{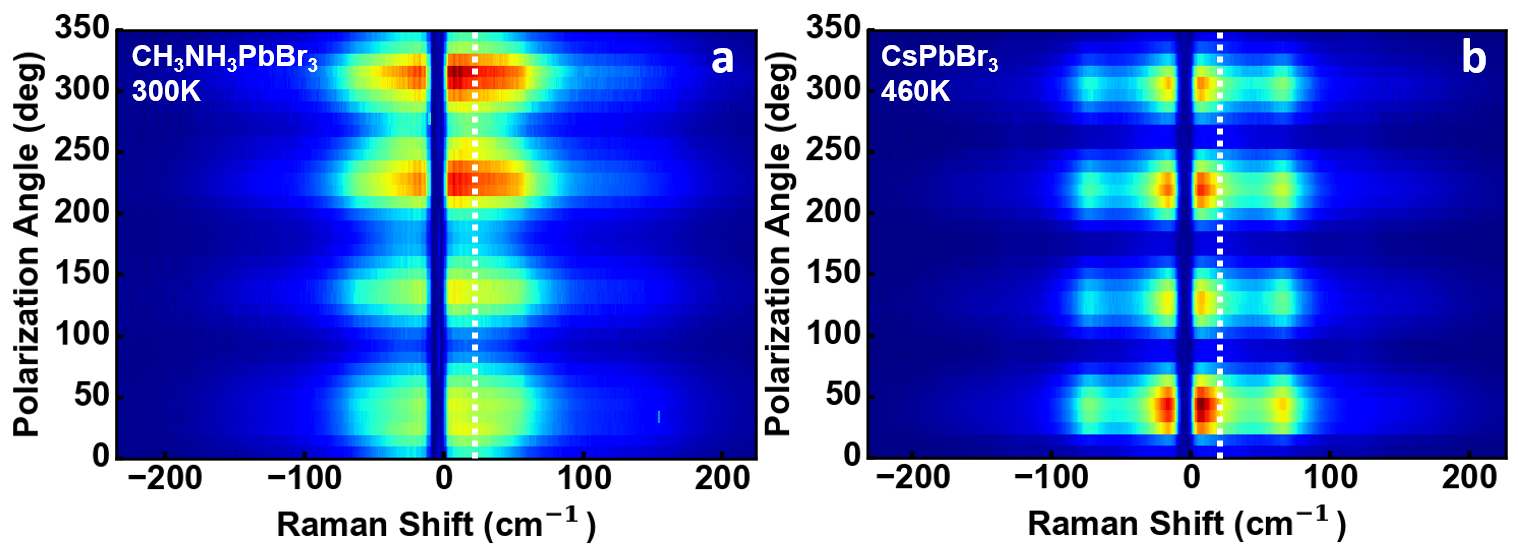}
\caption{
Cross-polarized low frequency Raman spectra of hybrid and inorganic lead-halide perovskite crystals in the cubic phase, (a) \MABrws (at 300 K) and b) \Csws (at 460 K), showing strong polarization angle dependence with 90$^\circ$ period in both materials. 
The zero of polarization angle is arbitrarily defined.
The white dotted line indicates a Raman shift of 20~\wn.  The corresponding modulation of the Raman intensity is shown in Fig. S3 in the supplemental material.
}
\label{fig:cross_pol}
\end{figure}

To learn more about the thermal fluctuations that give rise to the central peak, we measured the angular dependence of cross-polarized Raman spectra~\cite{Mizoguchi1989,Ribeiro2015} of both crystals in the cubic phase (Figure \ref{fig:cross_pol}a and b). In this experiment, record scattered light polarized perpendicular to the incident light, with incident and scattered light propagating along the direction of the [100] crystal axis. We repeat this measurement after rotating the linear polarization of the incident light (using a half wave-plate) with respect to the crystal surface. In perfectly ordered crystals, the angular modulation of the intensity of scattered light probes the symmetry of an observed mode~\cite{Mizoguchi1989}. In contrast, we probe a central peak that originates from anharmonic, disordered motion. Therefore, the  strong and clear angular modulation of the central peak, observed in both \MABrws (Fig.~\ref{fig:cross_pol}a) and \Csws (Fig.~\ref{fig:cross_pol}b) is surprising. It indicates that the disordered motion, giving rise to the central peak is highly anisotropic. Even more surprising is that both crystals exhibit similar modulation. This indicates that, although the motion of CH$_3$NH$_3^+$ and Cs$^+$ are expected to be very different, the modulation of the local polar fluctuations possess similar features for both crystals and must therfore be determined by their common lead-bromide octahedral network. Additionally, though the central peak of both crystals exhibits angular modulation, contrasting Figure \ref{fig:cross_pol}a with \ref{fig:cross_pol}b does reveal the effect of the CH$_3$NH$_3^+$ rotation. While \Csws exhibits a clear and well-defined minimum in the intensity, \MABrws exhibits a diffuse modulation minimum (see supplemental aterial for the modulation profile). This non-zero minimum reflects additional isotropic disorder that is induced by CH$_3$NH$_3^+$ rotation, similar to the motion of liquids where the Raman central peak is independent of excitation polarization angle.\cite{Walrafen1986} 

To better understand the motions that give rise to the central peak, we performed first-principles MD simulations of \Csws at 80 K and 500 K and calculated Raman spectra from the polarizability autocorrelation function of the resulting trajectories~\cite{Thomas2013}. Within this theoretical framework, the connection between Raman central peaks and dynamic local polar order has been established~\cite{Aksenov87p393,Vugmeister99p8602}. This can be seen by writing the Raman intensity ($I$) as \cite{Vugmeister99p8602}

\begin{multline}\label{pppp}
I^{\mu \nu}(\omega) \propto \\
 \int \, dq' d\omega' \,
\langle P^{\mu}(r,t) P^{\mu}(0,0) \rangle_{q',\, \omega'} \,
\langle P^{\nu}(r,t) P^{\nu}(0,0) \rangle_{-q',\, \omega-\omega'}
\end{multline}

\begin{figure}
\centering
\includegraphics[scale=0.3]{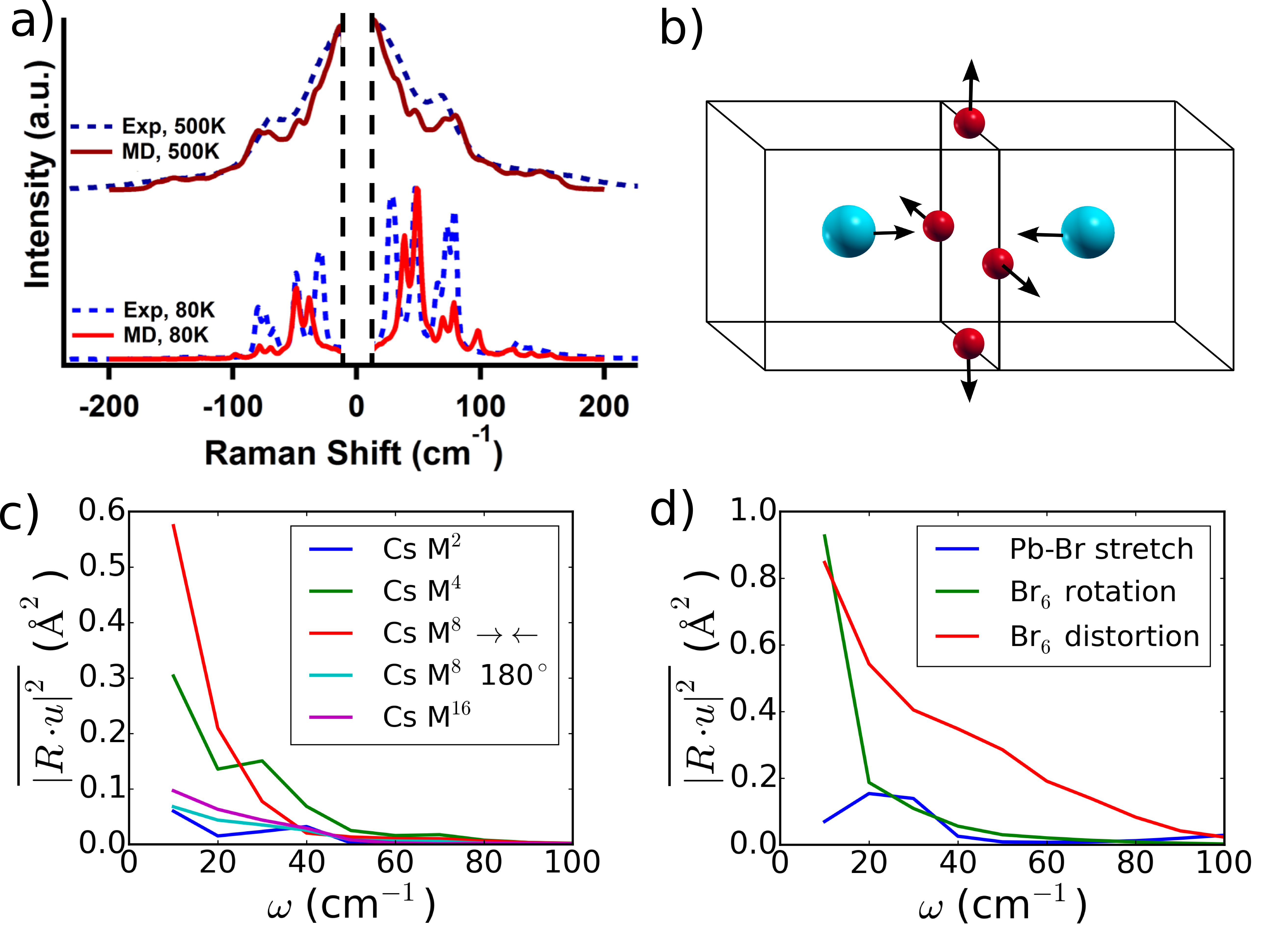}
\caption{
a) Comparison between MD (solid line) and experimental (dashed line) Raman spectra of CsPbBr$_3$ at $T=$80 K (bottom) and $T=$500 K (top).
b) Schematic representation of head-to-head Cs motion, coupled with motion of proximal Br, which is responsible for the Raman central peak in CsPbBr$_3$. Atoms are colored according to: Cs (cyan), Br (red). 
c) Projections of frequency-filtered MD trajectories to different Cs modes, categorized by multipole moment (see supplemental material for details). Weights of projections onto different Cs modes are given as a function of frequency of motion. 
d) Projections of frequency-filtered MD trajectories to different Br modes. Weights of projections onto different Br modes are given as a function of frequency of motion. 
}
\label{fig:theory}
\end{figure}

\noindent Here, $\omega$ is Raman frequency, $q$ is wave vector, $\langle P^{\mu}(r,t) P^{\mu}(0,0) \rangle$ are polarization-polarization correlation functions, where $\langle\cdot\rangle$ denotes an average over start times, and $\mu, \nu$ are cartesian components of $P$. A central peak occurs when the line shapes of the polarization-polarization correlation functions are peaked near $\omega=0$, that is, when there are quasi-static temporal correlations occurring over a range of time scales. Using polarizabilities obtained from density functional theory, we calculate Raman spectra that are in good agreement with experiment, showing a strong central peak at 500 K (Fig.~\ref{fig:theory}a).   

To obtain insight into the nature of fluctuations giving rise to this central peak, we examine our MD trajectories at 500 K. Root-mean-square displacements (Cs: 0.85 \AA, Pb: 0.36 \AA, Br: 0.73 \AA), as well as typical trajectories (supplemental material), show that Cs is the most mobile atomic species in this material, with Br displaying large fluctuations as well, while Pb is mostly static in comparison. Examination of MD trajectories indicates that instantaneous structures in the cubic phase are characterized by large off-center displacements of the Cs$^+$ cation as well as tilting and distortions of PbBr$_6$ octahedra. The angular distribution function of off-center Cs atoms shows that Cs is mainly displaced along the $\langle100\rangle$ directions (supplemental material).

The conditions necessary for a central peak are not only that the structure display large thermal fluctuations, but also that there be incipient polar order, with the stochastic appearance and disappearance of polar motifs. Such temporal behavior gives rise to vibrational response at a continuum of frequencies and a frequency-domain polarization correlation functions and Raman spectra peaked at $\omega=0$ (Eq.~\ref{pppp}). In contrast, a purely harmonic or weakly anharmonic material will not have spectral weight near $\omega=0$. In CsPbBr$_3$, we have found that the polar fluctuations giving rise to the central peak consist primarily of  head-to-head motions of Cs atoms, coupled with perpendicular, outward motion of the proximal Br atoms (Fig.~\ref{fig:theory}b). We present evidence for this in the mode-decompositions of frequency-filtered MD trajectories (Fig.~\ref{fig:theory}c). We filter our MD trajectories into frequency intervals of 10 cm$^{-1}$ using a band-pass filter, and project these filtered trajectories onto the different Cs modes allowed in our 2$\times$2$\times$2 simulation cell. We find particularly high weight for the modes $M^8$ ($\rightarrow\leftarrow$), which consist of head-to-head Cs motion (see supplemental material for mode labels), with increasing weight for lower frequencies. A similar analysis for the Br modes shows that motions of Br perpendicular to Pb-Br-Pb bonds dominate the low frequency spectral response (modes labelled as Br$_6$ rotations and distortions in Fig.~\ref{fig:theory}d). Linear combinations of these modes give rise to concerted outward motion of Br shown in Fig.~\ref{fig:theory}b.

While antiferroelectric modes have been linked to Raman central peaks in other systems~\cite{Tagantsev13p2229}, the head-to-head  motion presented above is a unique aspect of CsPbBr$_3$. Such motion is usually thought to be energetically unfavorable due to the opposing dipoles, as may be expected from the prevalence of 180$^\circ$ domain walls over head-to-head domain walls. However, in CsPbBr$_3$, the head-to-head Cs motion is stabilized by a cooperative motion of Br as Cs approaches one of the faces of the cubo-octahedral cage. This face, as defined by the positions of the 4 Br on it, tends to expand as this happens. It is energetically favorable for two neighboring Cs to move towards the same face, resulting in head-to-head motion, instead of the two Cs atoms separately causing expansions of two different cubo-octahedron faces. At these short distances, Pauli repulsion interactions dominate Coulomb effects in determining low energy modes. This process is illustrated in animations of the frequency-filtered trajectories {and confirmed with further projective analysis of the MD trajectory (see supplemental material)}.   

The picture {of} head-to-head motion of Cs atoms is consistent with the $\pi/2$ period of Raman intensity modulations in our cross-polarized measurements. Furthermore, the strong coupling between Cs and Br fluctuations revealed by the MD simulations is consistent with the pre-resonant Raman effect in CsPbBr$_3$, since electronic transitions involve valence band states with strong Br character. The similarities between our CsPbBr$_3$ and \MABrws pre-resonant Raman spectra suggest a similar picture in CH$_3$NH$_3$PbBr$_3$, with dipolar fluctuations driven by \MAws, but strongly coupled to Br motions which experience pre-resonant Raman. 
{Preliminary MD calculations of} \MABrws {indicate that Br motion does indeed contribute strongly to the central peak, while contributions from} \MAws {motion includes both rotational and displacive components (supplemental material).}

Central peaks have been calculated~\cite{Grinberg09p197601} and observed in other systems, such as relaxor materials~\cite{Siny97p7962,Jiang02p184301,Ko07p082903,Malinovsky10p1231,Shabbir04p167}, oxide perovskites including BTO~\cite{Malinovsky13p124,Ko08p102905} and niobates~\cite{Sokoloff1988,Sokoloff90p2398,Quittet76p5068,Kania1986,Bouziane2003}, and vitreous silica~\cite{Winterling1975,Buchenau1986}. However, there are differences in the local polar fluctuations of the halide and oxide perovskites arising from structural differences in these systems. Halide perovskites have half the formal charges of the oxide perovskites ({\em e.g}.,\ Cs$^+$, Pb$^{2+}$, Br$^{-}$ {\em vs}. \ Ba$^{2+}$, Ti$^{4+}$, O$^{2-}$ ), which makes Coulomb and ionic effects much less important in the halides~\cite{Edwardson89p9738}. The larger lattice spacing favors A-site fluctuations in the halide perovskites, while covalence effects due to lone pairs (Pb$^{2+}$, Bi$^{2+}$) favor B-site fluctuations in the oxides. In the oxides, covalency favors dipole alignment due to overbonding effects~\cite{Cockayne99p12542}.
On the other hand, Pauli repulsion tends to result in antiferroelectric fluctuations in the halide perovskites via the mechanism discussed above. Thus, the central peaks in the oxides and halides arise from different underlying dynamics.

In conclusion, we compare the low-frequency Raman scattering of \MABrws and \Csws crystals in each of their structural phases and show that both systems exhibit significant and surprisingly similar local polar thermal fluctuations in their cubic and tetragonal phases. Thus, the presence of a dipolar organic  cation is not a necessary condition for dynamic disorder in the halide perovskites, nor is rotational cation motion. 
Using MD simulations, we have shown  that  head-to-head Cs motion coupled to Br face expansion drives dynamic disorder in CsPbBr$_3$. In the halide perovskites, the weak bonds between the  A-site cation and PbX$_3$, together with the stronger bonds within the PbX$_3$ lattice, produce motion on multiple time scales, which is manifested in the observed diffuse central peak superimposed on the usual sharper features in the Raman spectra.

Structural fluctuations are known to be prevalent in the halide perovskites. Here we precisely identify which vibrational motifs have large amplitude at moderate temperature. 
{In CsPbBr$_3$, halide motion, coupled to A-site translation, is shown to be dominant. CH$_3$NH$_3$PbBr$_3$ shows analogous behavior, with halide motion, CH$_3$NH$_3$ translation, and CH$_3$NH$_3$ rotation all contributing to the central peak.}
Coupling of electronic excitations to these particular motions may deepen understanding of temporal band edge fluctuations~\cite{Motta2015}, carrier localization~\cite{Quarti2015}, and dynamical Rashba spin splitting~\cite{Zheng2015,Etienne16p1638}.

\subsection*{Acknowledgments}
This work was supported by the U.S. Department of Energy, Office of Science, Office of Basic Energy Sciences, with funding at Columbia University through the Energy Frontier Research Center under Grant No. DE-SC0001085 and at SLAC National Accelerator Laboratory through the AMOS program within the Chemical Sciences, Geosciences, and Biosciences Division. O.Y. acknowledges funding by the FP7 People program under the project Marie Curie IOF-622653. M.A.P. acknowledges support of the Brazilian Agencies CNPq and Fapemig.
L. Z. T acknowledges support of the Office of Naval Research, under grant number N00014-14-1-0761. F. Z. acknowledges support of the DOE Office of BES, under grant number DE-FG02-07ER46431. A. M. R. acknowledges support of the Office of Naval Research, under grant number N00014-12-1-1033. L.K. and D.A.E. were supported by a research grant from Dana and Yossie Hollander, in the framework of the Weizmann Institute of Science (WIS) Alternative Sustainable
Energy Research Initiative, AERI. D.A.E. further acknowledges financial support by the Austrian Science Fund (FWF): J3608-N20. MGK acknowledges funding from DOE-NNSA grant and thanks National Nuclear Security Administration for grant DE-NA0002522. We thank Shi Liu for the valuable discussions.

\bibliography{DDLHPC_2016_PRL}

\end{document}



\title{Supplemental Material - Local polar fluctuations in lead halide perovskite crystals}

\author{Omer Yaffe}
\thanks{These authors contributed equally}
\affiliation{Department of Chemistry, Columbia University, New York, NY 10027, USA}
\author{Yinsheng Guo}
\thanks{These authors contributed equally}
\affiliation{Department of Chemistry, Columbia University, New York, NY 10027, USA}
\author{Liang Z. Tan}
\affiliation{Department of Chemistry, University of Pennsylvania, Philadelphia, Pennsylvania 19104, USA}
\author{David A. Egger}
\affiliation{Department of Materials and Interfaces, Weizmann Institute of Science, Rehovoth 76100, Israel}
\author{Trevor Hull}
\affiliation{Department of Chemistry, Columbia University, New York, NY 10027, USA}
\author{Constantinos C. Stoumpos}
\affiliation{Materials Science Division, Argonne National Laboratory, Argonne, Illinois 60439, USA}
\author{Fan Zheng}
\affiliation{Department of Chemistry, University of Pennsylvania, Philadelphia, Pennsylvania 19104, USA}
\author{Tony F. Heinz}
\affiliation{Department of Applied Physics, Stanford University, Stanford, CA 94305, USA}
\affiliation{SLAC National Accelerator Laboratory, Menlo Park, California 94025, United States}
\author{Leeor Kronik}
\affiliation{Department of Materials and Interfaces, Weizmann Institute of Science, Rehovoth 76100, Israel}
\author{Mercouri G. Kanatzidis}
\affiliation{Materials Science Division, Argonne National Laboratory, Argonne, Illinois 60439, USA}
\affiliation{Department of Chemistry, Northwestern University, Evanston, Illinois 60208, USA} 
\author{Jonathan S Owen}
\affiliation{Department of Chemistry, Columbia University, New York, NY 10027, USA}
\author{Andrew M. Rappe}
\affiliation{Department of Chemistry, University of Pennsylvania, Philadelphia, Pennsylvania 19104, USA}
\author{Marcos A. Pimenta}
\affiliation{Department of Chemistry, Columbia University, New York, NY 10027, USA}
\affiliation{Departamento de Fisica, Universidade Federal de Minas Gerais, 30123-970 Belo Horizonte, Brazil}
\author{Louis E. Brus}
\email{leb26@columbia.edu}
\affiliation{Department of Chemistry, Columbia University, New York, NY 10027, USA}




\pacs{Valid PACS appear here}
\keywords{Dynamic disorder; Perovskites; Low frequency Raman scattering}
\maketitle

\subsection*{Experimental and Computational details}

\small{
\textit{\MABrws crystal growth.} Orange colored single crystals of \MABrws crystals were grown by diffusion of isopropyl alcohol vapor into a 1 M solution of 1:1 lead bromide (98\%, Sigma-Aldrich) and methylammonium bromide in N,N-dimethylformamide. Methylammonium bromide was synthesized from aqueous methylamine solution (40\% w/w, Alfa Aesar) and hydrobromic acid (48\%, Acros).

\textit{\Csws and \CsClws crystal growth.} \Csws and \CsClws were synthesized in a solid-state process described in Ref. \cite{Stoumpos2013a}. Orange, transparent crystals were grown via Bridgman method where \Csws powder was finely ground and placed in a fused silica tube. The tube is brought to a 10$^{-4}$ mbar vacuum and flame-sealed. The ampoule is attached to a clock mechanism and is slowly lowered into a 3-zone vertical tube furnace with a temperature gradient of 10$^{\circ}$C/mm. The dropping speed was varied between 3-30 mm/h. Small colorless cubic \CsClws crystals were grown similarly.

\textit{Low-frequency Raman scattering measurements.} Micro, low-frequency Raman measurements were taken using a custom-built setup with a 632.8 nm helium-neon laser. A dichroic 90/10 volume holographic grating (VHG) beamsplitter filter (NoiseBlockTM, Ondax, Inc.) redirected the laser towards the sample in an Oxford instruments microscopy cryostat (Microstat HiRes2 pumped to 10$^{-6}$ mBar), where a 40X/0.6 NA objective lens focuses the laser onto the sample \hl{(both spot size diameter and depth of field are {$\approx$2} $\mu$m)} and collects the back-scattered light. The 90/10 beamsplitter then rejects 90\% of the Rayleigh HeNe scatter back towards the laser while transmitting most of the Raman shifted signals. Two ultra-narrowband VHG notch filters (SureBlockTM, Ondax, Inc.), each having optical density $>$4.0, then further attenuate the collected Rayleigh scattered light while transmitting the Raman signals with an estimated system transmission efficiency of $>$80\%. The Raman signal passed through a spatial filter and into a single stage spectrometer with an 1800 grooves/mm diffraction grating (2 cm$^{-1}$ resolution) and liquid nitrogen cooled CCD array detector. For angular dependence of the polarized Raman scattering and cross-polarized  measurements,  linear polarizers were added before and after the sample, and a 
half-wave plate was used to rotate the initial polarization angle of the laser.
}

\textit{DFT and MD calculations.} 
First-principles MD simulations of orthorhombic CsPbBr$_3$ were performed using a pre-optimized supercell containing 4 formula units of CsPbBr$_3$, a $(6\times4\times6)$ $k$-point grid, and a planewave kinetic-energy cutoff of 250 eV. For cubic CsPbBr$_3$, we have used a supercell with 8 formula units of CsPbBr$_3$ and a $(4\times4\times4)$ $k$-point grid. The total energy in the DFT-part of the MD simulations is converged up to $10^{-6}$ eV per unit cell, {and we do not account for spin-orbit coupling as this was found to not affect the structural properties of lead-halide perovskites.\cite{Egger2014} Optimizations of the lattice parameters of CsPbBr$_3$ were performed with the same settings, except that we have used a larger planewave kinetic-energy cutoff of 400 eV to compute accurate stresses.\cite{Egger2014}}
The first principles MD on CsPbBr$_3$ at 80K and 500K were performed with the VASP code,\cite{Kresse1996} using the PBE functional\cite{Perdew1996} and the Tkatchenko-Scheffler scheme\cite{Tkatchenko2009} to account for dispersive interactions. We equilibrated the systems using Born-Oppenheimer MD runs of 5 ps and a Nos\'{e}-Hoover thermostat. The standard deviation per atom for the last ps of these equilibrations were {$\approx$} $10^{-4}$ eV and {$\approx$} 1 K for the (conserved) total energy and temperature, respectively. The MD trajectories to obtain the velocity autocorrelations were 40 ps long, with a 10 fs time step. We verified that at this simulation length, the velocity autocorrelation was converged. To obtain the Raman spectra, we calculate the polarizability at time intervals of 100 fs using density functional perturbation theory (DFPT) \cite{Gonze1997}. Raman spectra were obtained from the autocorrelation function of the polarizability \cite{Thomas2013}. To mitigate numerical noise at low frequencies, we employed zero-padding in the computation of the autocorrelation function, and calculated the autocorrelation of the deviation of the polarizability: $\langle (\alpha(t+\tau)-\bar{\alpha})(\alpha(t)-\bar{\alpha}\rangle _t$, where $\alpha(t)$ is the polarizability at time $t$ and $\bar{\alpha}$ is its time-average. The polarizability and phonon properties were computed using Quantum Espresso\cite{Giannozzi2009}  {with the PBE functional.}\cite{Perdew1996}  {Norm-conserving, designed non-local pseudopotentials}\cite{Ramer1999,Rappe1990}  {were used for all the atoms.}

\subsection*{Phase sequence of \MABrws and \Csws}
\begin{table}[H]
\caption{\label{tab:phasesequence}Phase sequence and temperature range for \MABrws\cite{Poglitsch1987,Mashiyama2007} and \Csws\cite{Stoumpos2013a,Hirotsu1974} from x-ray diffraction data. Note that the transition temperatures of \Csws are much higher than those of \MABr. }
\begin{ruledtabular}
\begin{tabular}{llll}
\hline
Crystal&\multicolumn{1}{c}{Temp. (K)} &\multicolumn{1}{c}{Crystal system}&Space Group  \\ \hline
\multicolumn{1}{l}{\multirow{4}{*}{\MABr}}&\multicolumn{1}{l}{$<$144.5}&\multicolumn{1}{l}{Orthorhombic}&\multicolumn{1}{l}{Pbnm / P222$_1$}\\
\multicolumn{1}{l}{} &\multicolumn{1}{l}{149.5-155.1}&\multicolumn{1}{l}{Tetragonal}&\multicolumn{1}{l}{P 4/mmm}\\
\multicolumn{1}{l}{} &\multicolumn{1}{l}{155-237}&\multicolumn{1}{l}{Tetragonal}&\multicolumn{1}{l}{I4/mcm}\\
\multicolumn{1}{l}{}&\multicolumn{1}{l}{$>$237}&\multicolumn{1}{l}{Cubic}&\multicolumn{1}{l}{Pm3m}\\
\hline
\multicolumn{1}{l}{\multirow{3}{*}{\Cs}}&\multicolumn{1}{l}{$<$370}&\multicolumn{1}{l}{Orthorhombic}&\multicolumn{1}{l}{Pbnm}\\
\multicolumn{1}{l}{} &\multicolumn{1}{l}{370-420}&\multicolumn{1}{l}{Tetragonal}&\multicolumn{1}{l}{P 4/mbm}\\
\multicolumn{1}{l}{}&\multicolumn{1}{l}{$>$420}&\multicolumn{1}{l}{Cubic}&\multicolumn{1}{l}{Pm$\bar{3}$m}\\
\hline
\end{tabular}
\end{ruledtabular}
\vspace{1cm}
\end{table}

\subsection*{Fitting of the Raman scattering spectra with one Debye relaxation and multiple damped Lorentz oscillator components}
The quasi-elastic scattering spectral continuum, namely the central peak, is a characteristic feature of relaxational dynamics in many materials, observed by X-ray~\cite{Krakauer1999} and neutron scattering~\cite{Mashiyama2007}, as well as Raman scattering~\cite{Fontana1988,Fontana1990,Bouziane2003,Sokoloff1988,Hushur2005}. 

Following the treatment in ref~\onlinecite{Fontana1990,Bouziane2003}, we use the imaginary part of the Debye relaxation model for the central peak and the imaginary part of the damped Lorentz oscillator model for the Raman peaks. A Lorentzian lineshape approximation is often used in Raman spectroscopy, when the damping is much smaller than the resonance frequency. However, since in lead-halide perovskites the resonance frequency is very low, this approximation no longer holds.  We therefore employ the damped Lorentz oscillator model to fit the discrete Raman modes. The experimentally measured Raman scattering spectrum is expressed as eq. \ref{eq:exp},
\begin{equation}
I_{exp}(\nu)=c_{\text{B-E}}(\nu)
\left(
\frac{ c_0 \left| \nu \right| \gamma_0 }{\nu^2 +\gamma_0^2}+
\sum_{i=1}^{n} \frac{ c_i \left| \nu \right| \gamma_i^3} {\nu^2 \gamma_i^2+(\nu^2 -\nu_i^2)^2}
\right)
\label{eq:exp}
\end{equation}
where $\nu$ is the spectral shift, $\nu_i$  is the resonance energy of the damped Lorentz oscillators, $\gamma_0$ and $\gamma_i$ are the damping coefficients of the Debye relaxation and Lorentz oscillators, $c_0$ and $c_i$ are unitless fitting parameters for the intensities of the Debye relaxor and Lorentz oscillator components respectively. The spectral shift $\nu$, the parameters $\nu_i$, $\gamma_i$ and $\gamma_0$ are in wavenumber units. The Debye relaxation time $\tau_r$ is related to $\gamma_0$ as $\tau_r=\frac{1}{2\pi c \gamma_0}$, where c is the speed of light 

The prefactor $c_{\text{B-E}}(\nu)$ accounts for the thermal population from the Bose-Einstein distribution. The Stokes signal is proportional to 1+n, and the anti-Stokes signal is proportional to n, with n being the Bose-Einstein distribution. The Bose-Einstein prefactor is written as eq. \ref{eq:BE}
\begin{equation}
c_{\text{B-E}}(\nu)= 
\begin{cases}
n(\nu)+1 & \nu \geq 0 \\
n(\left|\nu \right| ) & \nu < 0
\end{cases}
\label{eq:BE}
\end{equation}

To model the scattering data, we first normalize the experimental Raman scattering data with the product of Bose-Einstein population factor and a spectral shift factor from both Debye and Lorentz terms. This normalization accounts for the universal thermal population's contribution to the low frequency spectral intensity, and allows access to the intrinsic vibrational and relaxational dynamics. We remove the spectral artifacts from the notch filter around 0 \wn, and fit both the normalized anti-Stokes and Stokes scattering simultaneously with equation \ref{eq:fitting} using fitting parameters $c_0, \gamma_0, c_i, \nu_i, \gamma_i$. Numerical fitting was carried out in the Python program with leastsqbound-scipy package\cite{Helmus}, and in Igor Pro 6. 
\begin{equation}
I_{n}(\nu)=\frac{I_{exp}(\nu)}{c_{\text{B-E}}(\nu) \left| \nu \right|}= 
\frac{ c_0 \gamma_0 }{\nu^2 +\gamma_0^2}+
\sum_{i=1}^{n} \frac{ c_i \gamma_i^3 }{\nu^2 \gamma_i^2+(\nu^2 -\nu_i^2)^2}
\label{eq:fitting}
\end{equation}

\begin{figure}
\centering
\includegraphics[scale=0.7]{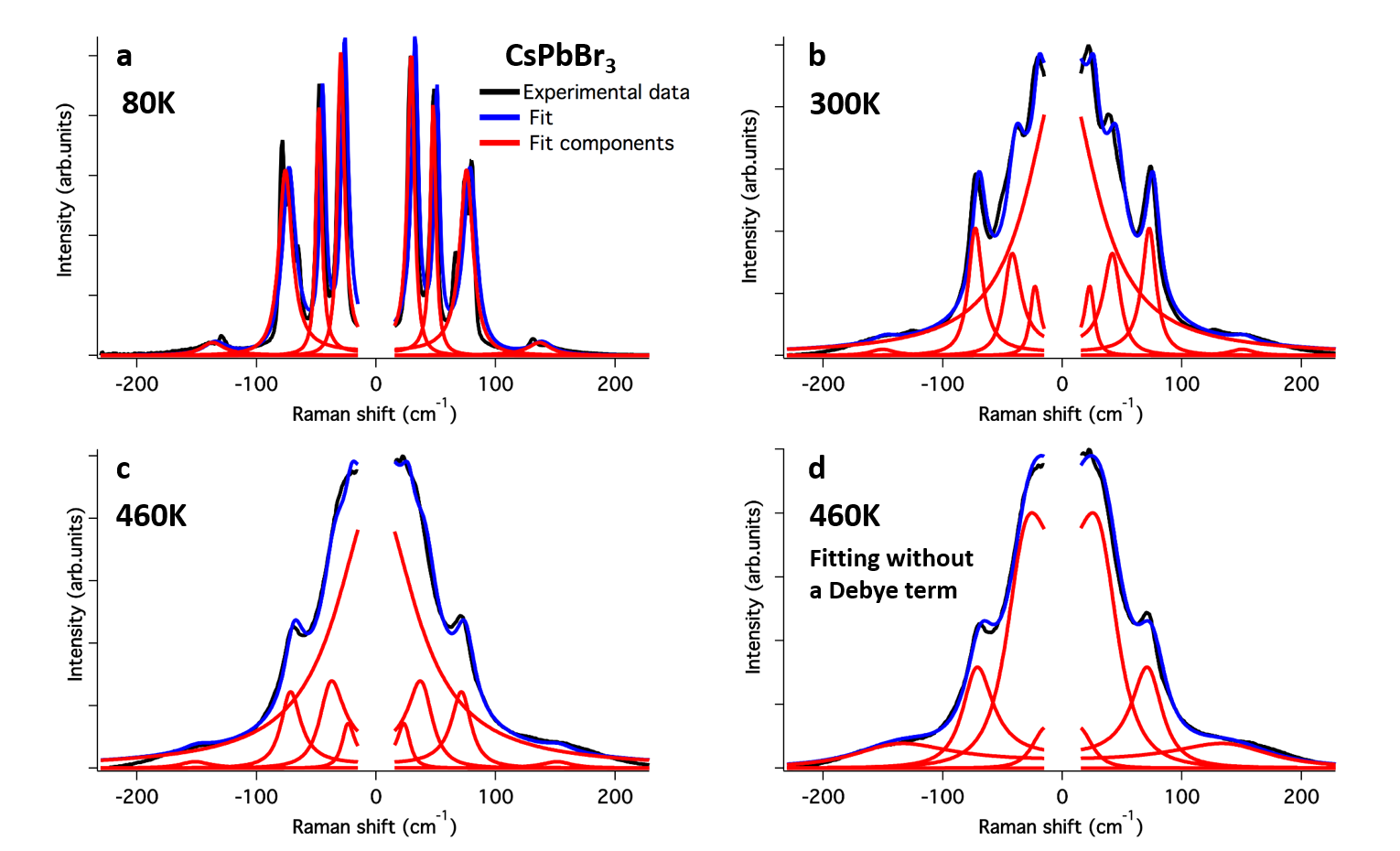}
\caption{
Examples of fitting low frequency Raman spectra in \Cs.  A) 80~K. Individual Raman modes are grouped and represented by 4 Lorentz terms. The ratio of the Debye term area relative to the whole spectrum is $1.9\times10^{-4}$. B) 300~K. Raman modes broaden and merge, the background spectral continuum rises. The relative weight of the Debye term is 0.62. C) 460~K fit with Debye and Lorentz terms. The relative weight of the Debye term is 0.66. D) 460~K fit with only Lorentz terms. Good quality fit can also be obtained without a Debye term. The Raman modes at ca. 24 \wnws and 100 \wnws are heavily overdamped, producing an asymmetrical lineshape and serving as the spectral continuum. Black lines are experimental data. Blue lines are the overall fit to the data. Red lines are individual components separately plotted.}
\label{fig:fitting}
\end{figure}

Next we discuss possible limitations of our model in fitting the low frequency Raman scattering data. 
The spectral intensities of low-lying Lorentz oscillators overlap with the Debye relaxor. This spectral overlap introduces ambiguity in fitting the central peak. Therefore, we set physical constraints on the Lorentz terms based on extrapolation from the low temperature Raman spectra. In Figure~\ref{fig:fitting} we present our fitting procedure, using \Csws as an example.

\begin{figure}
\centering
\includegraphics[scale=0.7]{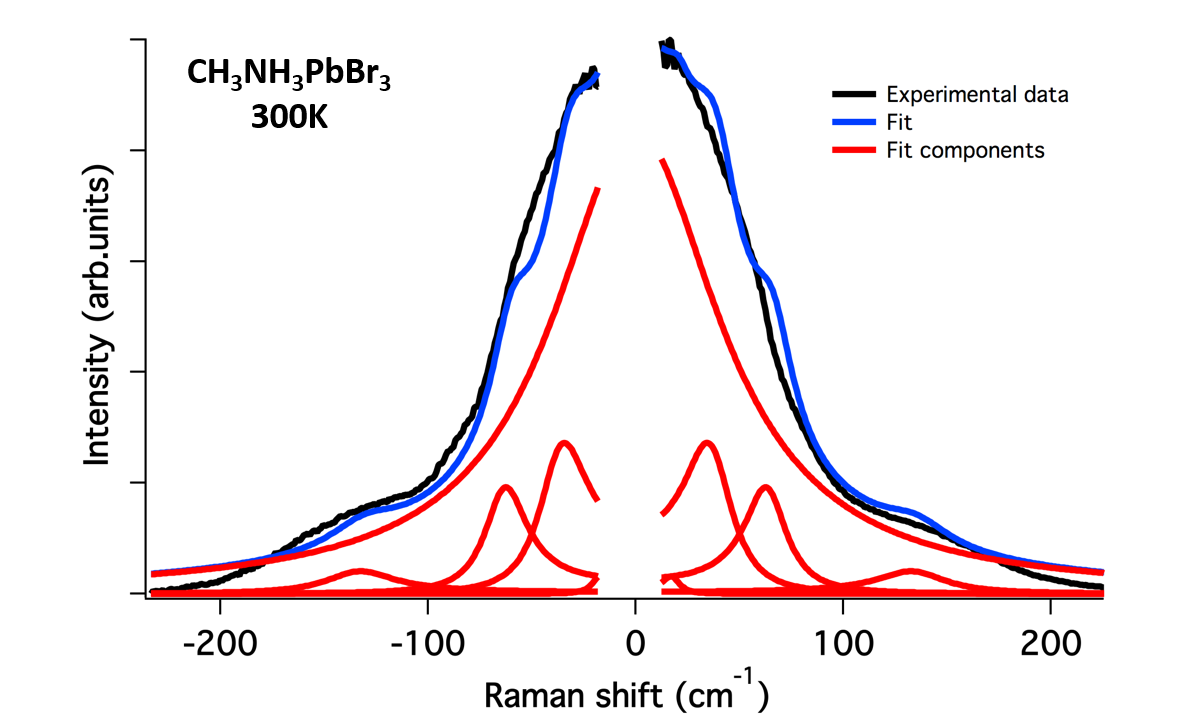}
\caption{
Examples of fitting low frequency Raman spectra in cubic phase \MABr. Black lines are experimental data. Blue lines are the overall fit to the data. Red lines are individual components separately plotted.}
\label{fig:fitting_MA}
\end{figure}

 The \Csws Raman modes are grouped into 4 regions (ca. 20~\wn, 40~\wn, 70~\wn, and 140~\wn), each represented by a Lorentz term. We extrapolate the observed softening and broadening of each Raman mode from the temperature dependent Raman spectra at low temperatures. At high temperatures, the position of each Lorentz term is constrained to be within 5 \wnws of the value expected from extrapolation. The width of each Lorentz term is constrained to be no larger than the value expected from extrapolation. In figure~\ref{fig:fitting}a, \ref{fig:fitting}b and \ref{fig:fitting}c, we show the evolution of fittings at 80~K, 300~K and 460~K respectively. The fitting parameters are shown in table~\ref{table_fitting}. As temperature increases, the spectral continuum rises, and the relative weight of the Debye component increases and becomes dominant. Figure~\ref{fig:fitting}c shows an example of a Debye relaxation time of $\approx$130 fs for \Csws in the cubic phase. Figure~\ref{fig:fitting_MA} shows a Debye relaxation time of $\approx$ 110~fs for \MABrws in the cubic phase.
 \begin{table}
\caption{\label{table_fitting} Fitting parameters of \Csws and \MABrws low frequency Raman spectra}
\begin{ruledtabular}
\begin{tabular}{llllllll}
\hline
Crystal&\multicolumn{1}{c}{Temperature (K)} &\multicolumn{1}{c}{Debye width (\wn)} &\multicolumn{1}{c}{Lorentz position (\wn)} &\multicolumn{1}{c}{Lorentz width (\wn)} &Notes  \\ 
\hline
\multicolumn{1}{l}{\multirow{4}{*}{\Cs}}&\multicolumn{1}{c}{80$^*$}&\multicolumn{1}{c}{0.1}&\multicolumn{1}{l}{29.8, 48.0, 76.4, 137.7} &\multicolumn{1}{l}{7.0, 5.6, 13.3, 23.3} &\multicolumn{1}{l}{orthorhombic phase}\\
\multicolumn{1}{l}{} &\multicolumn{1}{c}{300}&\multicolumn{1}{c}{34.2}&\multicolumn{1}{l}{24.0, 43.5, 73.5, 151.4} &\multicolumn{1}{l}{9.9, 17.0, 14.2, 23.7}&\multicolumn{1}{l}{tetragonal phase}\\
\multicolumn{1}{l}{} &\multicolumn{1}{c}{460}&\multicolumn{1}{c}{39.8}&\multicolumn{1}{l}{25.0, 41.0, 72.8, 153.0}&\multicolumn{1}{l}{12.8, 25.0, 20.0, 30.0} &\multicolumn{1}{l}{cubic phase}\\
\multicolumn{1}{l}{}&\multicolumn{1}{c}{460}&\multicolumn{1}{c}{-}&\multicolumn{1}{l}{24.0, 41.5, 74.3, 150.9}&\multicolumn{1}{l}{30.6, 46.5, 31.2, 100.0}&\multicolumn{1}{l}{no Debye term}\\
\hline
\multicolumn{1}{l}{\MABr}&\multicolumn{1}{c}{300}&\multicolumn{1}{c}{48.9}&\multicolumn{1}{l}{18.4, 40.0, 65.0, 135.7}&\multicolumn{1}{l}{10.0, 28.8, 25.0, 41.2} &\multicolumn{1}{l}{cubic phase}\\
\hline
\label{tab:fitting_parametrs}
\end{tabular}
\end{ruledtabular}
\vspace{1cm}
{$^*$Note: 80~K was experimentally measured temperature. The anti-Stokes and Stokes scattering intensity was normalized to be symmetrical with a Bose-Einstein population factor at 110~K. Such adjustment was not necessary for other spectra at higher temperatures.}
\label{table_fitting}
\end{table}

Alternatively, we remove the Debye term and keep the multiple Lorentz oscillators without constraints from low tempearture, producing the fit shown in Figure~\ref{fig:fitting}d.
The Raman modes are further softened and overdamped to produce a good overall fit containing the spectral continuum.
This all-Lorentz  fitting model does not assume or account for newly activated spectral intensity.
However, our MD simulations reveal that the motions that give rise to the low frequency scattering have weak restoring forces as in the Debye model. 
Moreover, relaxational motion occures in many perovskite-structured materials, in which the Debye relaxation model was used~\cite{Poglitsch1987, Fontana1988, Fontana1990, Bouziane2003, Sokoloff1988, Hushur2005}.
This difficulty of distinguishing between an order-disorder relaxational motion and a softened normal mode has been discussed for classes of perovskites~\cite{Fontana1988, Fontana1990, Zalar2003}.
Both perspectives point to a picture of overdamped, heavily anharmonic lattice motion at low frequency as the essential characteristic of dynamic disorder.

\clearpage

\subsection*{Cubic phase (high temperature) angular dependence cross section of perpendicular polarized, low frequency Raman spectra}

\begin{figure}[H]
\centering
\includegraphics[scale=0.5]{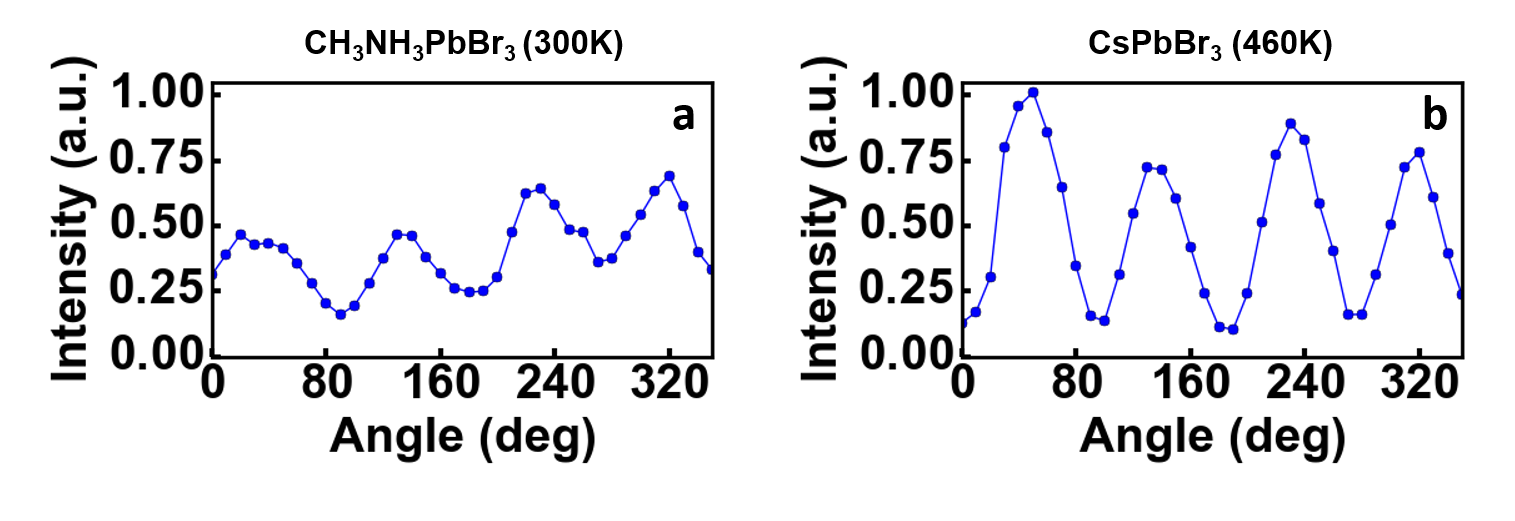}
\caption{
Angular dependence cross section of perpendicular polarized, low frequency Raman sepctra of \MABrws at 300 K (a) and \Csws at 460 K (b). These profiles correspond to the dotted line in figure 2 of the main text
}
\label{fig:angle_dep_300K}
\end{figure}

\subsection*{Density functional theory and molecular dynamics - Peak assignment of the low-temperature, orthorhombic phase }
\begin{figure}[H]
\centering
\includegraphics[scale=0.5]{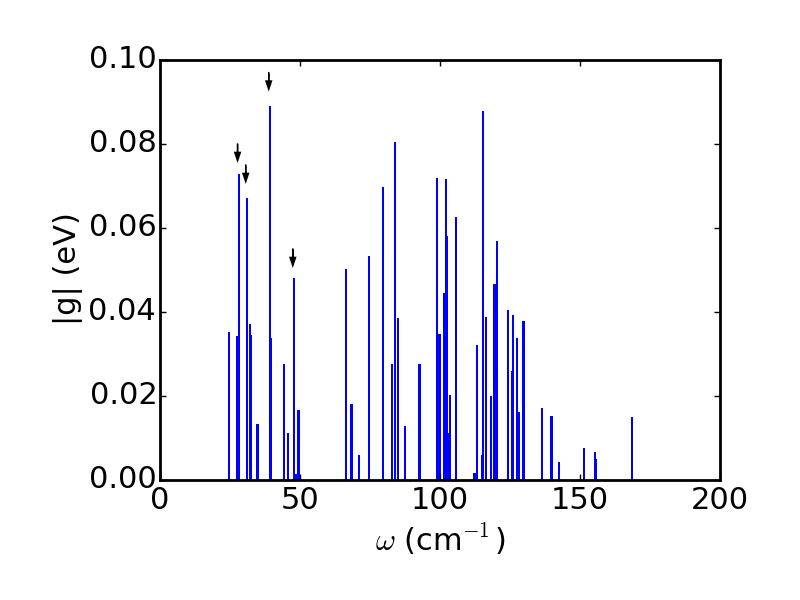} 
\caption{Electron-phonon matrix elements for CsPbBr$_3$ in the Pbnm phase. Arrows denote the modes depicted in \ref{fig:modes}.
}
\label{fig:eph}
\end{figure}

\begin{figure}[H]
\centering
\includegraphics[scale=0.15]{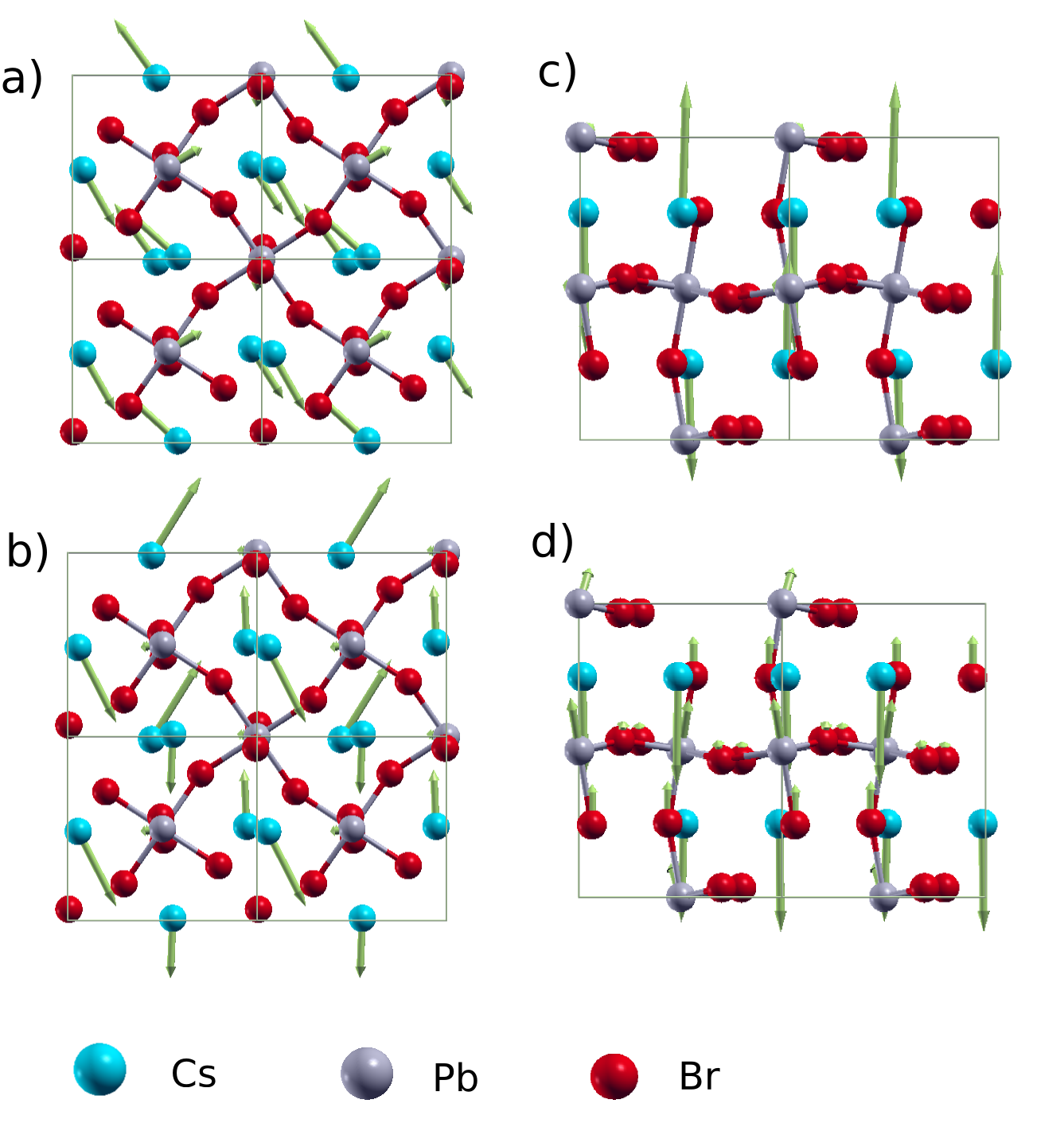} 
\caption{Selected phonon modes with strong electron-phonon couplings, calculated from DFT, for CsPbBr$_3$ in the Pbnm phase. The frequencies of the modes are: a) 27.70 cm$^{-1}$, b) 30.61 cm$^{-1}$, c) 38.83 cm$^{-1}$, d) 48.01 cm$^{-1}$. Panels a) and b) are displayed in the [001] direction, while panels c) and d) are displayed in the [110] direction. Green arrows indicate motion.
}
\label{fig:modes}
\end{figure}

We calculate phonon modes and electron-phonon coupling matrix elements of P${bnm}$ phase \Csws as described in the experimental section. Figure \ref{fig:eph} shows the electron-phonon coupling strengths of low frequency modes at zero wave-vector. Selected modes having significant Cs and Pb motion are illustrated in figure \ref{fig:modes}. Modes (b) and (d) are A$_g$ symmetry with frequencies in good agreement with the strong peaks observed in 80 K Raman spectra. In Fig. S2 harmonic frequencies are given for low amplitude motion; these modes may show anharmonic behavior at higher amplitudes.

\subsection*{Density functional theory and molecular dynamics - Anharmonic, disordered motion}

\begin{figure}
\centering
\includegraphics[scale=0.3]{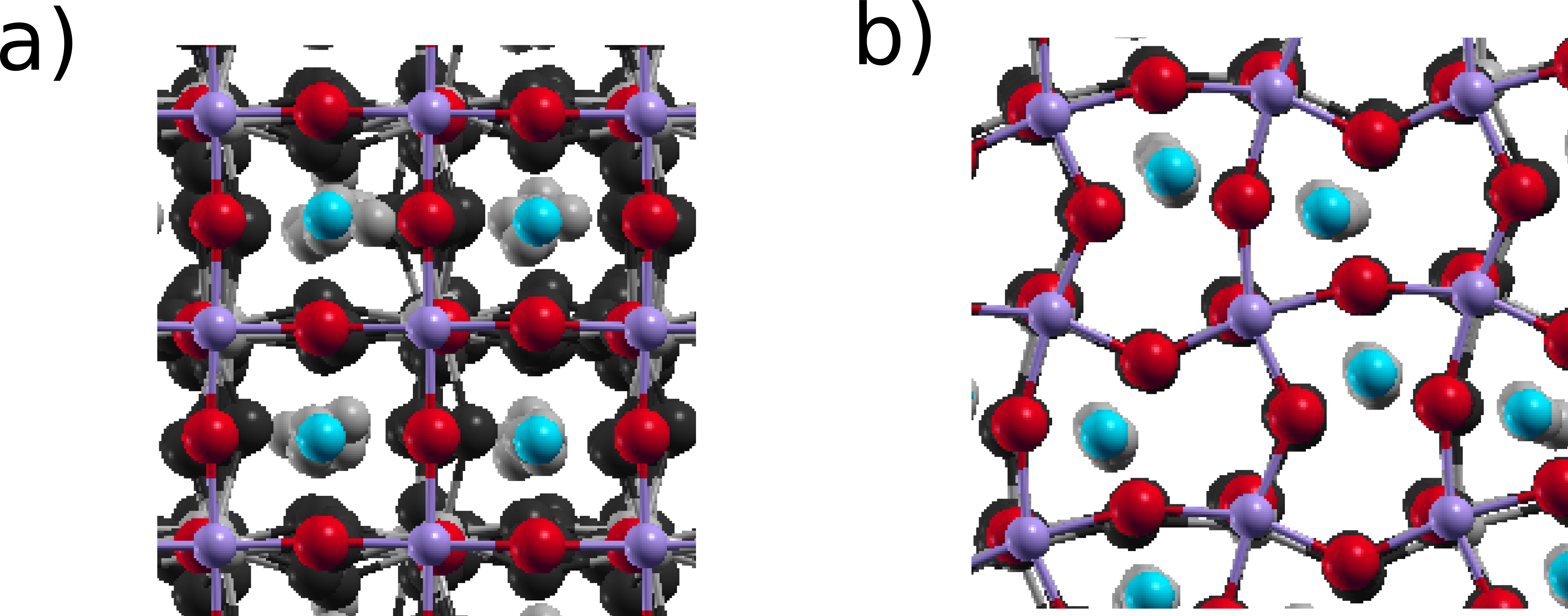} 
\caption{
Schematic representation of instantaneous and averaged CsPbBr$_3$ atomic structure obtained in molecular dynamics simulations. In color: averaged positions of atoms over entire MD trajectory, in grayscale: instantaneous structures selected at random over entire MD trajectory. In each panel, 10 instantaneous images are plotted. a) Cubic phase, at 500 K. b) Orthorhombic phase, at 80K.  Atoms are colored according to: Cs (cyan), Pb (purple), Br (red). 
}
\label{fig:snapshots}
\end{figure}

\begin{figure}
\centering
\includegraphics[scale=0.3]{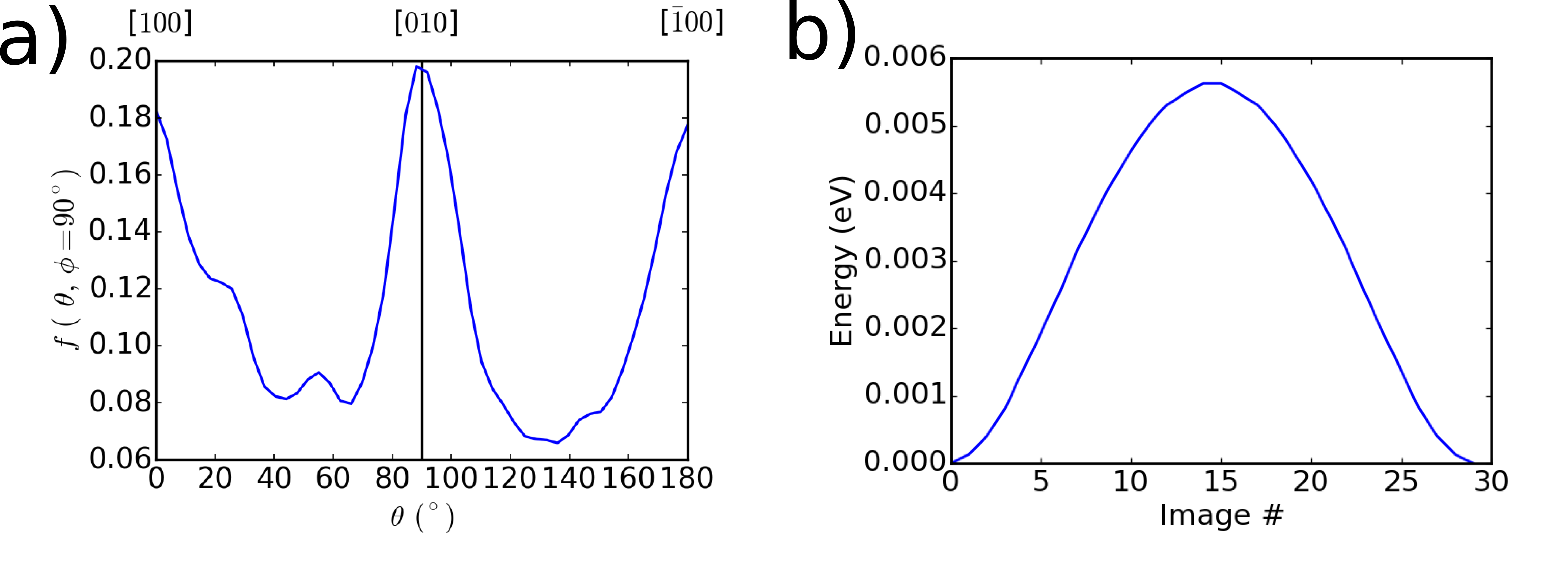} 
\caption{
a) Angular probability distribution function of Cs off-center displacements at 500 K, calculated along the equatorial plane (containing [100] and [010] directions). b) Energy barrier between off-centered Cs sites for the averaged 80 K structure. Image 0 and image 30 correspond to Cs off-centered positions of the averaged 80 K structure. Energies are plotted for interpolated structures between these two off-centered sites.
}
\label{fig:csprob}
\end{figure}

We show instantaneous and averaged structures obtained from MD simulations in Fig.~\ref{fig:snapshots}. The averaged 500 K structures display cubic symmetry, while the instantaneous 500 K structures display large deviations from cubic symmetry, with Cs displacements and distortions of Pb-Br-Pb bonds. Analysis of Cs displacement distribution function shows that Cs is mostly displaced along the $\langle 100 \rangle$ directions (Fig.~\ref{fig:csprob}a). At 80 K, equilibrium positions of Cs atoms are already off-centered in the cubo-octahedral cages. They are displaced either in the [100] or [010] directions from the centers of cubo-octahedral cages. Density functional theory calculations show that the energy barrier between two off-centered Cs positions within a single cubo-octahedtal cage of the 80 K structure is 5.8 meV (Fig.~\ref{fig:csprob}b). In contrast, there is no energy barrier for Cs positions in the cubo-ocatahedral cage for the 500 K structure.


\begin{figure}
\centering
\includegraphics[height=5cm]{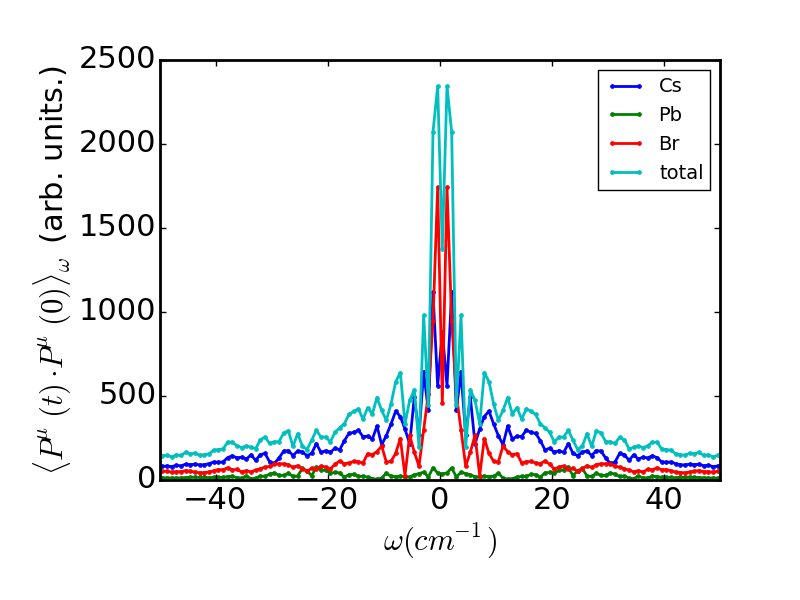}

\caption{Polarization-polarization correlation functions in the frequency domain, decomposed into atomic components. }
\label{fig:PPcorrelation}
\end{figure}

\begin{figure}
\centering
\includegraphics[width=0.8\textwidth]{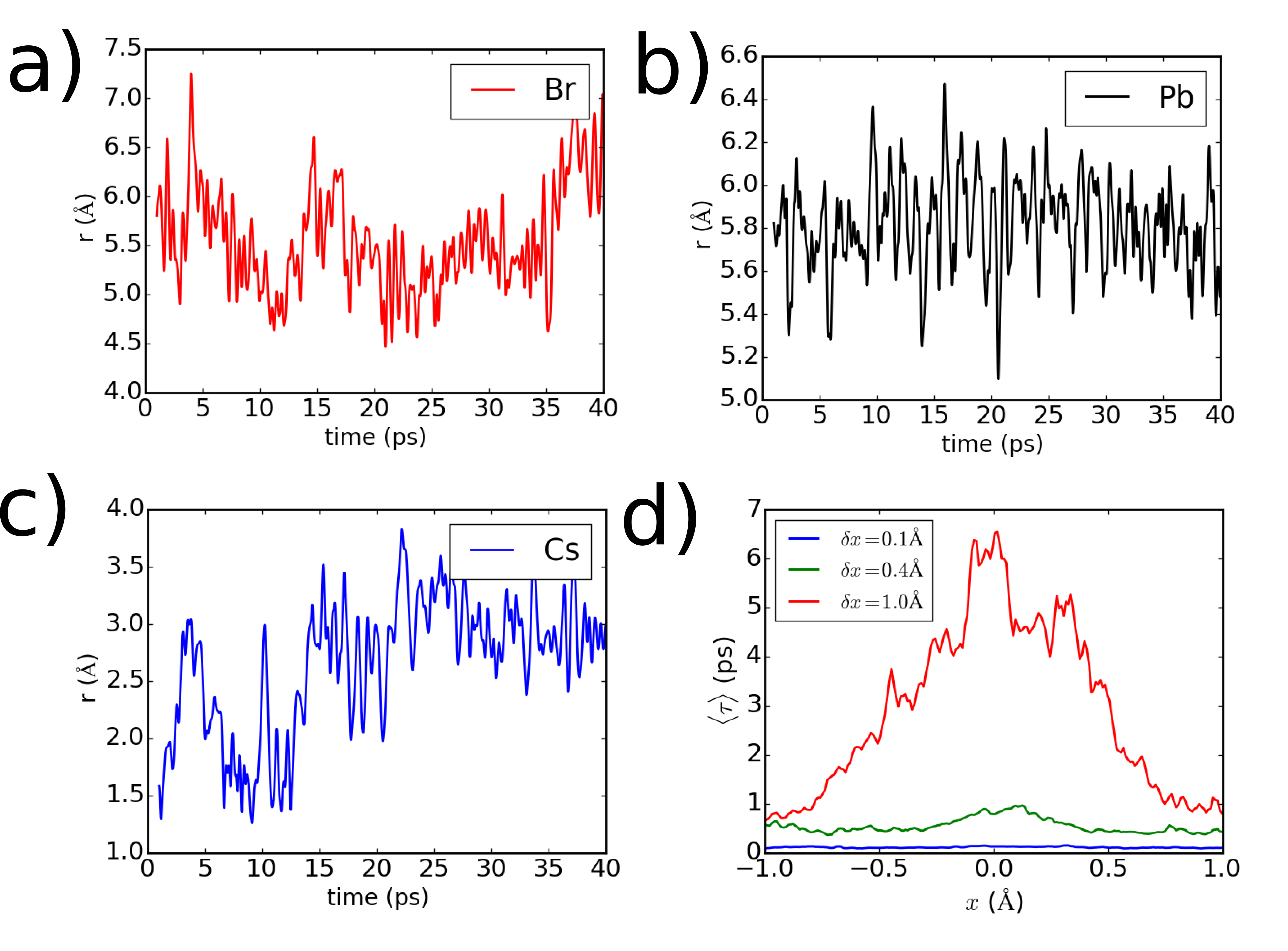}

\caption{Representative trajectories of Br (a), Pb (b), and Cs(c) atoms in at T=500K MD simulation. d) Dwell times (constructed from a running average over all trajectories) of Cs atoms as a function of position, for the 500 K MD trajectory. The dwell time is defined as the time taken
for a Cs atom to move \hl{a distance $\delta x$ from its original position. Here, we show dwell times for $\delta x=0.1,0.4,1.0$ \AA. As $\delta x$ increases beyond the length scale for Cs motion ($\sim 1.0$ \AA), the dwell times at the average position ($x=0$) increase substantially because Cs atoms will rarely deviate more than $\delta x$ from the average position.  } }
\label{fig:500ktraj}
\end{figure}

We calculate polarization autocorrelation functions from the MD trajectories using displacements of atoms in cage-centered coordinates, together with nominal charges for the ions (Fig.~\ref{fig:PPcorrelation}). The autocorrelation functions are defined as

\begin{equation}
\langle P^{\mu}(t) \cdot P^{\mu}(0) \rangle_\omega = \int dt' dt \,  P^{\mu}(t') \cdot P^{\mu}(t+t')\, e^{-i\omega t}
\end{equation}

\noindent For Cs displacements, we use $P^{Cs}=r_{Cs}-(1/12)\sum_{j=1}^{12} r_{Br,j}$. For Pb displacements, we use $P^{Pb}=2r_{Pb}-(2/6)\sum_{j=1}^{6} r_{Br,j}$. For Br displacements, we use $P^{Br}=r_{Br}-(1/8)\sum_{j=1}^{4} r_{Cs,j}-(1/4)\sum_{i=1}^{2} r_{Pb,i}$. Here, the $r_{Cs,j}$, $r_{Pb,j}$, $r_{Br,j}$ refer to the nearest neighbor Cs, Pb and Br atoms. This figure shows that Cs and Br contribute strongly to polarization fluctuations, but not Pb.

\begin{figure}[H]
\centering
\includegraphics[width=\textwidth]{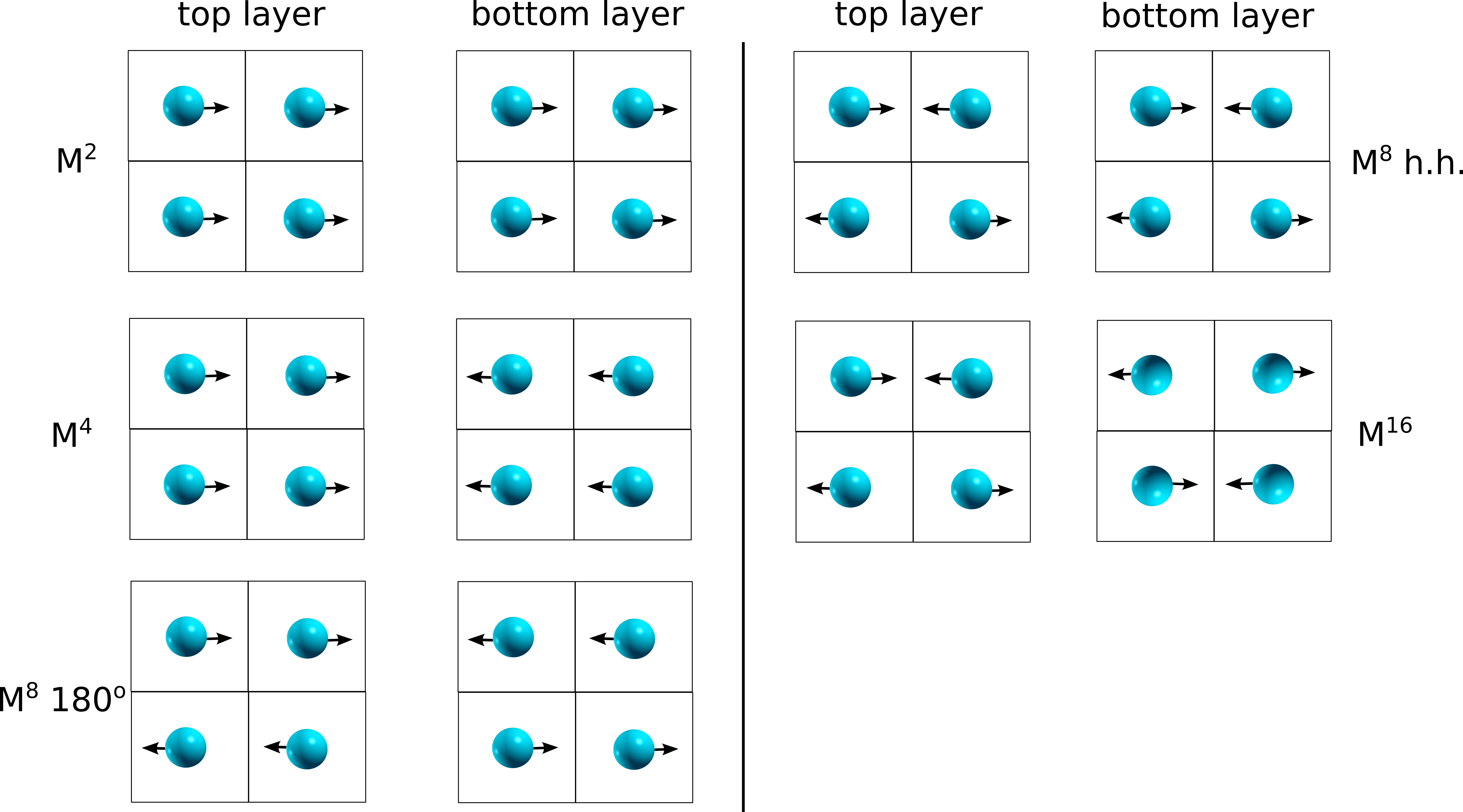}

\caption{Mode labels for Cs motion in a 2$\times$2$\times$2 simulation cell, classified according to multipole moment, used in Fig.~3c of the main text. }
\label{fig:csmodelabels}
\end{figure}

MD trajectories were frequency filtered according to

\begin{equation}\label{eq:filtered}
\vec{R}_{filtered}(t;\omega_1) = \mathcal{F}^{-1}\left[
\Theta(\omega-(\omega_1-\delta\omega))
\Theta(-\omega+(\omega_1+\delta\omega))
\mathcal{F}[\vec{R}(t)]\right]
\end{equation}

\noindent where $\vec{R}_{filtered}(t;\omega_1)$ is a vector of cartesian components of atomic positions of the filtered trajectory, and $\mathcal{F}$ denotes the Fourier transform. The filter is applied at target frequency $\omega_1$ with bandwidth $2\delta\omega$. These filtered trajectories were used to construct the Cs and Br projections in Fig.~3 of the main manuscript, with modes labelled according to Fig.~\ref{fig:csmodelabels}. An animation of a 500 K MD trajectory filtered at 10 cm$^{-1}$ is attached as a supplemental file. In this animation, Cs atoms are represented by blue circles. The area of each cubo-octahedron face separating the Cs atoms are drawn as rectangles. The color of the rectangles represents the area of this face, as defined by the 4 Br atoms on its vertices. Green indicates a bigger area than average, and red smaller. In this animation, each face defined by these Br atoms expands (green) as Cs on both sides moves towards it. This provides an illustration of the cooperative motion giving rise to the temporal antiferroelectric order which is responsible for the central peak. 

\hl{Projective analysis of the frequency filtered trajectories shown in the main text was done by projecting $\vec{R}_{filtered}$ in Eq.~\ref{eq:filtered} onto different modes ($\vec{u}$). That is, we compute the quantity
}

\begin{equation}
\lvert R\cdot u\rvert^2 = \frac{1}{N_t}\sum_t \lvert \vec{R}_{filtered}(t;\omega_1) \cdot \vec{u} \rvert^2
\end{equation}

\hl{
\noindent where the averaging was done over all time steps of the filtered trajectory, and for all filtered frequencies.
}

\begin{figure}[H]
\centering
\includegraphics[width=\textwidth]{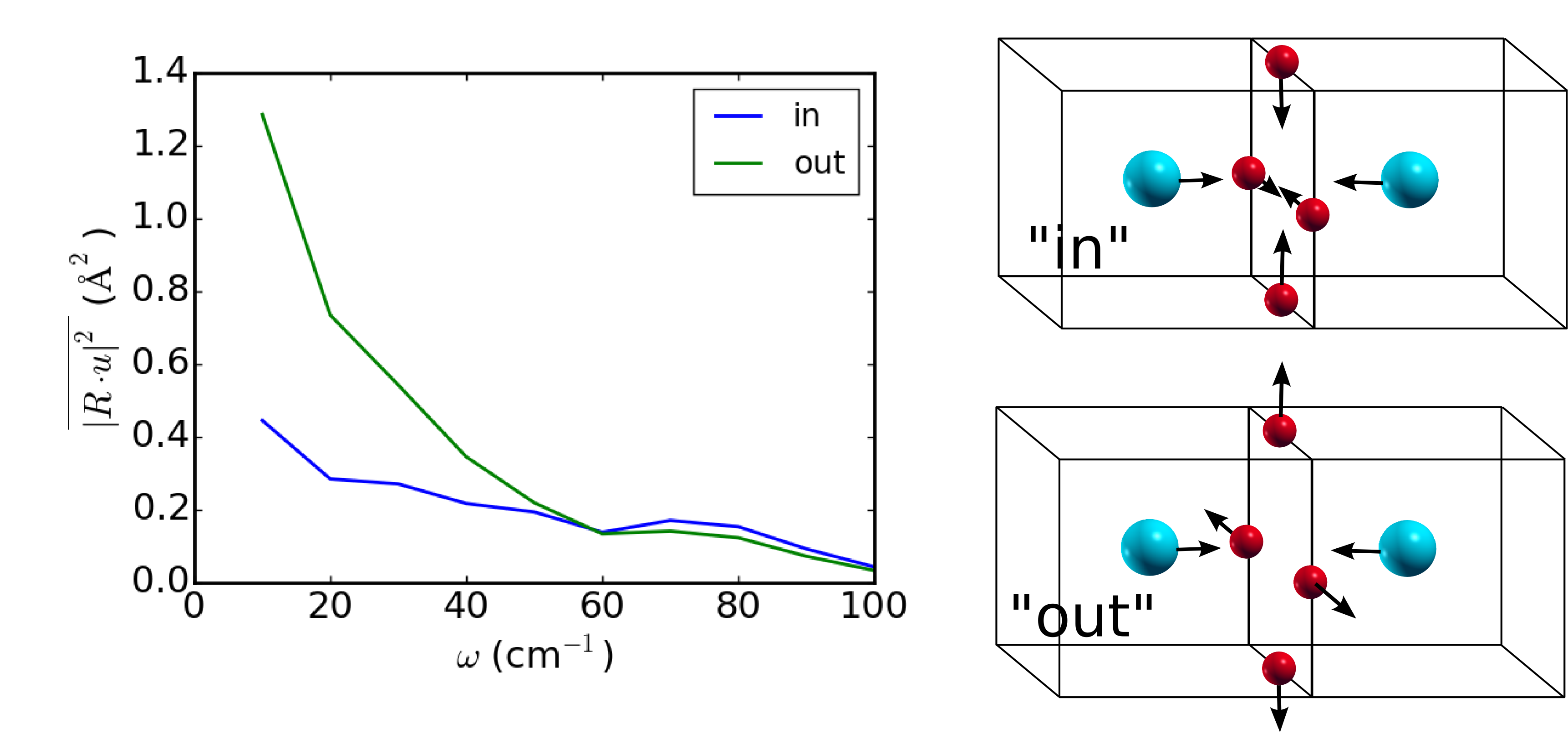}

\caption{\hl{Projections of frequency-filtered MD trajectories of CsPbBr$_3$ to two different Cs/Br coupled modes. The mode labelled "out" represents an expansion of the Br$_4$ window seen the animation above, while the mode labelled "in" represents a shrinkage of the Br$_4$ window instead. Weights of projections onto different Br modes are given as a function of frequency of motion. } }
\label{fig:csbrproj}
\end{figure}

\hl{
To quantify the magnitude of the Br part of the coupled Br/Cs head-to-head motion, we use projective analysis of the frequency filtered trajectories described above. In Fig.~\ref{fig:csbrproj}, we compare the coupled Cs head-to-head and Br$_4$ expansion seen in the animation above with another mode with the opposite sign of Br motion, corresponding to a shrinking of the Br$_4$ window instead. The larger weights of the former confirm that the coupling between Cs and Br motion as of the type shown in Fig.~3 of the main text.
}

\subsection*{Molecular dynamics simulations for CH$_3$NH$_3$PbBr$_3$}

\begin{figure}[H]
\centering
\includegraphics[width=0.5\textwidth]{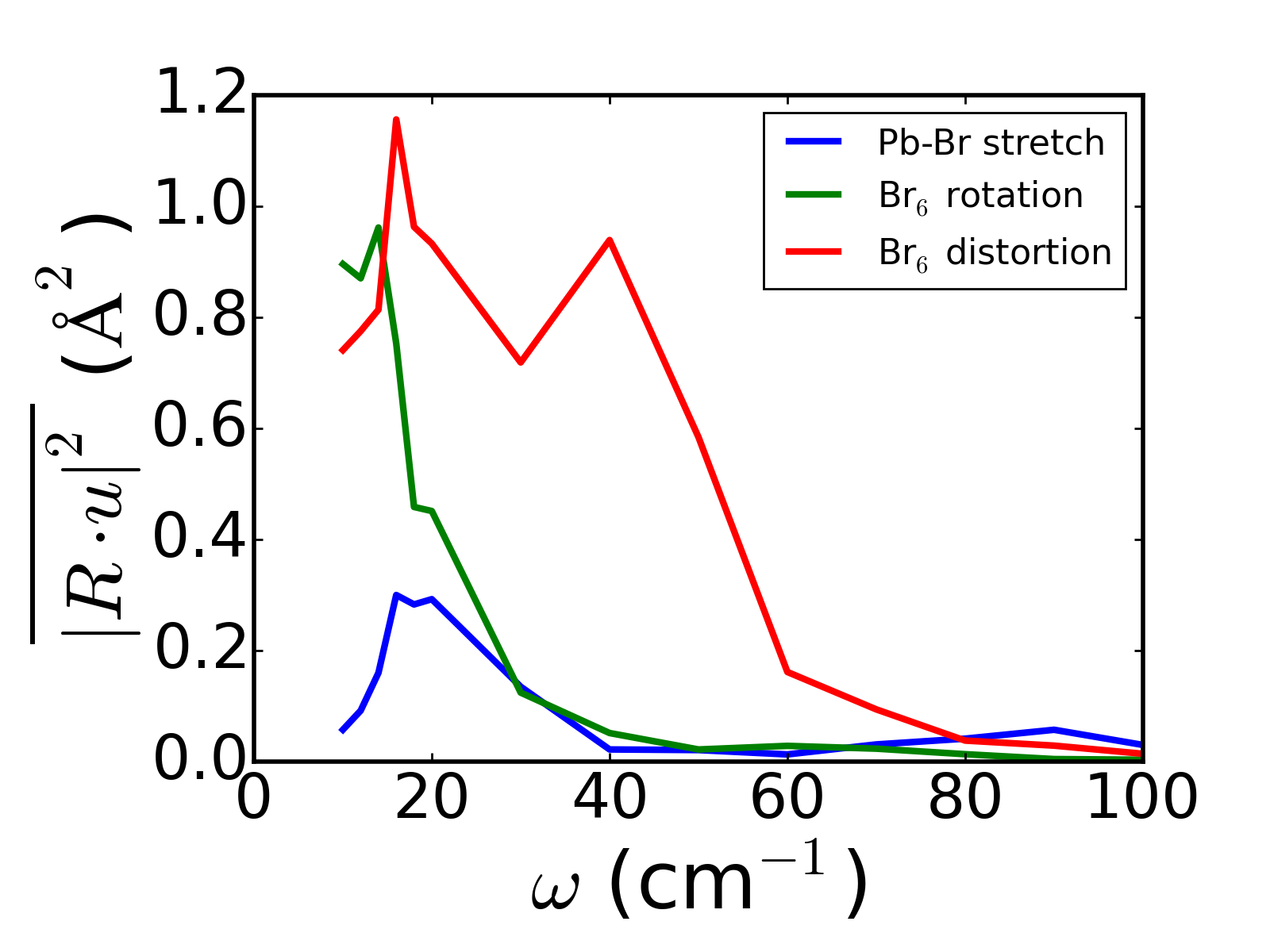}

\caption{\hl{Projections of frequency-filtered MD trajectories of CH$_3$NH$_3$PbBr$_3$ to different Br modes. Weights of projections onto different Br modes are given as a function of frequency of motion. }}
\label{fig:mabrproj}
\end{figure}

\begin{figure}[H]
\centering
\includegraphics[width=1.0\textwidth]{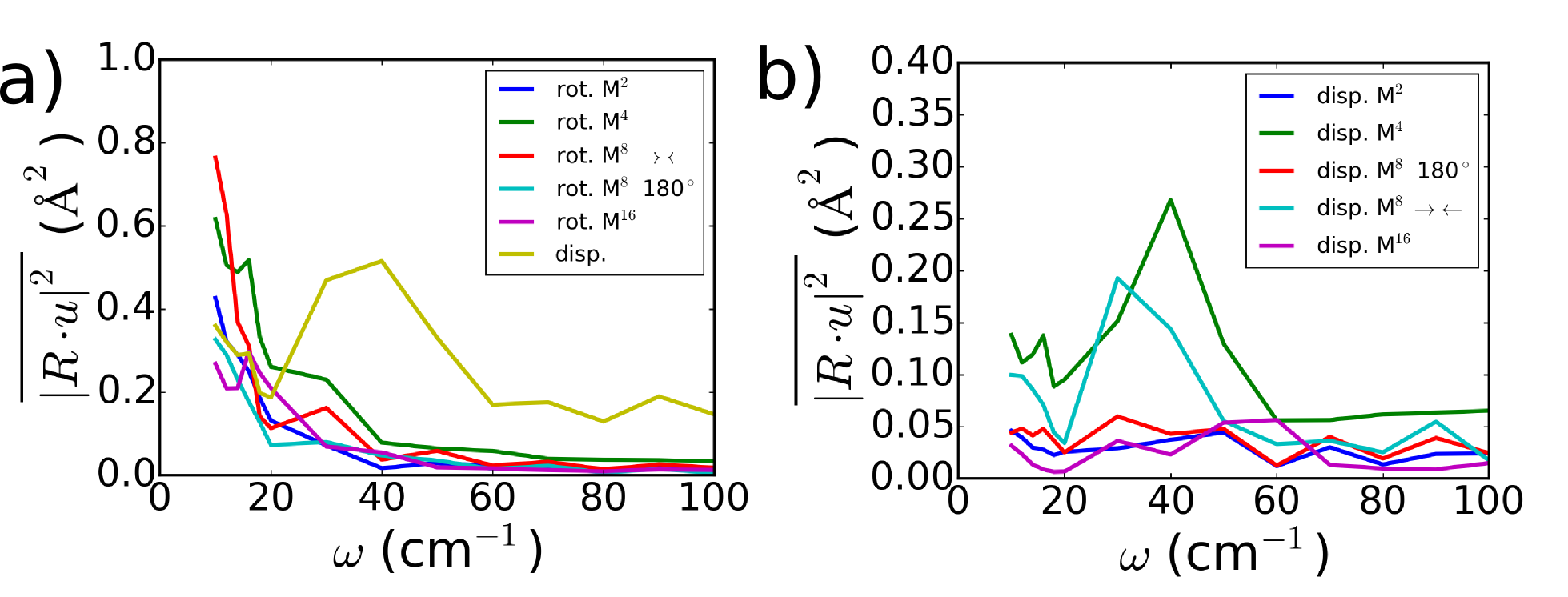}

\caption{\hl{Projections of frequency-filtered MD trajectories of CH$_3$NH$_3$PbBr$_3$ to different CH$_3$NH$_3$ modes. a) Comparison of  displacive (disp.) motion with various rotational (rot.) modes. Here, displacive motion refers to center-of-mass movement of CH$_3$NH$_3$, while rotational motion refers to changes in the orientation of CH$_3$NH$_3$. The mode labels used here refer to the same symmetries as in Fig.~\ref{fig:csmodelabels}. b) Decomposition of displacive motion into different modes. The mode labels used here refer to the same symmetries as in Fig.~\ref{fig:csmodelabels}.  Weights of projections onto different Br modes are given as a function of frequency of motion. }}
\label{fig:maproj}
\end{figure}

\hl{
We have performed and analyzed DFT-MD simulations of CH$_3$NH$_3$PbBr$_3$ using the same methodology as for CsPbBr$_3$, but with a time step of 1 fs. As this increases the computational costs by one order of magnitude compared to the MD for CsPbBr$_3$, the trajectory length of the CH$_3$NH$_3$PbBr$_3$ simulations was only 15 ps. Nevertheless, this allows for obtaining qualitative trends: The Br motion for CH$_3$NH$_3$PbBr$_3$ shows similar trends as for CsPbBr$_3$, with the dominance of Br displacements perpendicular to Pb-Br bonds (Fig.~\ref{fig:mabrproj}).  The presence of molecular dipoles in CH$_3$NH$_3$PbBr$_3$ introduces new molecular rotational modes. In Fig.~\ref{fig:maproj}, we compare the weight of CH$_3$NH$_3$ rotational modes with center-of-mass displacive motion of CH$_3$NH$_3$ as a function of frequency. While CH$_3$NH$_3$ rotational modes dominates at the lowest frequencies, there is a frequency range ($\sim40 cm^{-1}$) where displacive motion of CH$_3$NH$_3$ peaks, and has a higher weight than rotational motion. The characteristic frequency for CH$_3$NH$_3$ displacive motion is higher than that of Cs because CH$_3$NH$_3$ is lighter than Cs, and it occupies more of the cubo-octahedral PbBr cage. A further decomposition of CH$_3$NH$_3$ displacive motion into modes of different multipole moments shows that head-to-head motion of CH$_3$NH$_3$ accounts for a significant portion of the total CH$_3$NH$_3$ displacive motion. 
}

\par\bigskip 
\noindent

\bibliography{DDLHPC_2016_PRL}